\documentclass[aps,onecolumn,notitlepage,superscriptaddress]{revtex4-1}

\usepackage{bm}
\usepackage{graphicx}
\usepackage{latexsym}
\usepackage{revsymb}
\usepackage{amssymb}
\usepackage{amsmath}
\usepackage{mathtools}
\usepackage[colorlinks=true,citecolor=blue,linkcolor=blue]{hyperref}
\usepackage{bbding}
\usepackage[caption=false]{subfig} 
\usepackage{float}
\usepackage{subfloat}
\usepackage[section]{placeins}
\usepackage{csquotes}
\usepackage{relsize}
\usepackage{enumitem}
\usepackage{titlesec}
\usepackage{mfirstuc}
\usepackage[Symbol]{upgreek}
\usepackage{fancyhdr}
\usepackage{lastpage}
\usepackage{setspace}
\DeclareGraphicsExtensions{.pdf,.png,.jpg}

\LetLtxMacro{\OldSqrt}{\sqrt}
\newcommand{\ClosedSqrt}[1][\hphantom{3}]{\def\DHLindex{#1}\mathpalette\DHLhksqrt}
\makeatletter
    \newcommand*\bold@name{bold}
    \def\DHLhksqrt#1#2{%
        \setbox0=\hbox{$#1\OldSqrt{#2\,}$}\dimen0=\ht0\relax%
        \advance\dimen0-0.2\ht0\relax
        \setbox2=\hbox{\vrule height\ht0 depth -\dimen0}%
        {\hbox{$#1\expandafter\OldSqrt\expandafter[\DHLindex]{#2\,}$}
        \lower\ifx\math@version\bold@name0.6pt\else0.4pt\fi\box2}
    }
    \renewcommand*{\sqrt}[2][\ ]{\ClosedSqrt[\leftroot{-2}\uproot{1}#1]{#2}\kern0.1em} 
\makeatother

\newcommand{\dblinefill}{%
\leavevmode \cleaders \hbox to.2em{\hss=\hss}\hfill\kern-3pt
}

\renewcommand\vec{\mathbf}

\titleformat{\section}
  {\centering\Large\bfseries}{\thesection.}{1em}{}

\begin{document}


\title{ 
\dblinefill \\
\vspace{2mm}
Submitted to the Proceedings of the US Community Study \\ on the Future of Particle Physics (Snowmass 2021) \\
\vspace{2mm}
\dblinefill \\ 
\vspace{1.0cm}
\Large{PetaVolts per meter Plasmonics: Snowmass21 White Paper}
\vspace{3.0mm}
}

\author{Aakash A. Sahai}
\email[corresponding author: ~]{aakash.sahai@ucdenver.edu}
\affiliation{University of Colorado Denver, CO 80204}

\author{Mark Golkowski}
\affiliation{University of Colorado Denver, CO 80204}

\author{Stephen Gedney}
\affiliation{University of Colorado Denver, CO 80204}

\author{Thomas Katsouleas}
\affiliation{University of Connecticut, Storrs, CT}

\author{Gerard Andonian}
\affiliation{University of California, Los Angeles, CA}

\author{Glen White}
\affiliation{Stanford Linear Accelerator Center, Menlo Park, CA}

\author{Joachim Stohr}
\affiliation{Stanford Linear Accelerator Center, Menlo Park, CA}

\author{Patric Muggli}
\affiliation{Max-Planck-Institut fur Physik, Munich, Germany}

\author{Daniele Filipetto}
\affiliation{Lawrence Berkeley National Laboratory, Berkeley, CA}

\author{Frank Zimmermann}
\affiliation{European Center for Nuclear Research (CERN), Geneva, Switzerland}

\author{Toshiki Tajima}
\affiliation{University of California Irvine, CA}

\author{Gerard Mourou}
\affiliation{Ecole Polytechnique, Paris, France}

\author{Javier Resta-Lopez}
\affiliation{Institute of Materials Science, University of Valencia, Spain}

\begin{abstract}
\vspace{5.0mm}
\textbf{\abstractname.}
\setstretch{1.5}
Plasmonic modes offer the potential to achieve PetaVolts per meter fields, that would transform the current paradigm in collider development in addition to non-collider searches in fundamental physics. PetaVolts per meter plasmonics relies on collective oscillations of the free electron Fermi gas inherent in the conduction band of materials that have a suitable combination of constituent atoms and ionic lattice structure. As the conduction band free electron density, at equilibrium, can be as high as $\rm 10^{24}cm^{-3}$, electromagnetic fields of the order of $\rm 0.1 \sqrt{\rm n_0(10^{24}cm^{-3})} ~ PVm^{-1}$ can be sustained by plasmonic modes. Engineered materials not only allow highly tunable material properties but quite critically make it possible to overcome disruptive instabilities that dominate the interactions in bulk media. Due to rapid shielding by the free electron Fermi gas, dielectric effects are strongly suppressed. Because the ionic lattice, the corresponding electronic energy bands and the free electron gas are governed by quantum mechanical effects, comparisons with plasmas are merely notional. Based on this framework, it is critical to address various challenges that underlie PetaVolts per meter plasmonics including stable excitation of plasmonic modes while accounting for their effects on the ionic lattice and the electronic energy band structure over femtosecond timescales. We summarize the ongoing theoretical and experimental efforts as well as map out strategies for the future. Extreme plasmonic fields can shape the future by not only bringing tens of TeV to multi-PeV center-of-mass-energies within reach but also by opening novel pathways in non-collider HEP. In view of this promise, we invite the scientific community to help realize the immense potential of PV/m plasmonics and call for significant expansion of the US and international R\&D program.
\end{abstract}

\maketitle

\pagebreak
\tableofcontents

\pagebreak
\section*{\normalsize\MakeUppercase{Executive summary}}
\sectionmark{PetaVolts per meter Plasmonics: Executive summary}
\addcontentsline{toc}{section}{EXECUTIVE SUMMARY}

PetaVolts per meter acceleration and focusing gradients of plasmonic modes can both transform the paradigm of collider development as well as open a wide-range of non-collider studies of phenomena at the frontiers of high-energy physics (HEP). In the near-term, {\bf PetaVolts per meter plasmonics} will make it possible to directly access unprecedented fundamental tests of non-collider HEP. This non-collider paradigm effort described below will be pursued in parallel with the rather long-term goal of an integrated collider design using plasmonic accelerators.

The {\bf decadal goals} of the PV/m plasmonics effort, are as follows:
\begin{enumerate}[topsep=4pt, itemsep=0.3ex, partopsep=0.3ex, parsep=0.3ex]
 \item {\bf ``Opening the vacuum''} with ultra-strong plasmonic fields by approaching the ExaVolts per meter Schwinger field limit based on {\bf plasmonic nano-focusing} of the beam. Specifically, localized plasmonic fields that approach the Schwinger field increase the probability of spontaneous pair production directly off the vacuum, 
 \item {\bf ``Examining quantum gravity models''} with ultra-energetic gamma photons produced by a {\bf nano-wiggler} that wiggles the electrons of an ultrarelativistic non-collider beam ($\gamma_{\rm b}\simeq 10^{4-7}$) at the nanometric scale. Nanometric wiggling of the trajectories of ultrarelativistic electrons in plasmonic fields produces quasi-coherent GeV to TeV photons. It is theoretically predicted that the vacuum exhibits a non-trivial permittivity inversely proportional to the photon wavelength. The most appropriate probes for this model are sources that produce short bursts of ultra-energetic photons which are currently not available even at the lower end of the range above.
\end{enumerate}

PetaVolts per meter plasmonics relies on the unmatched density of the free electron Fermi gas at equilibrium. Because the free electron Fermi gas is a quantum mechanical entity inherent in the conduction band of conductive materials, several different {\bf quantum mechanical effects} work together in unison to make it the highest accessible equilibrium leptonic density, of the order of $\rm 10^{24} cm^{-3}$. Moreover, this dense Fermi electron gas is free to move in response to external excitations without the need for any special preparation of the conductive sample even at ambient pressure and temperature.

The electromagnetic fields of collective oscillations scale as the square root of the electron density of the media. Therefore, electromagnetic fields of the order of $\rm 0.1 \sqrt{\rm n_0(10^{24}cm^{-3})} ~ PVm^{-1}$ can be sustained by plasmonic modes which redefine the limits of high-field science and advanced accelerators.

As our PetaVolts per meter plasmonics initiative is a novel concept anchored in plasmonic physics, it is imperative to distinguish and disambiguate it from other existing techniques that are out of the scope of PV/m plasmonics. To this effect, the critical differences from ultra-dense plasmas and dielectric or insulating solids are summarily highlighted. While the density of gaseous plasmas is limited by handling of dense gasses, solid-state plasmas although ultra-dense are a major challenge for stability and consistency because they are ablated from solids.

The innovation of high-amplitude or nonlinear and relativistic collective surface crunch-in plasmonic oscillations of the free electron Fermi gas driven by ultrashort sources of particles has been theoretically modeled in three dimensions using collisionless particle-tracking methods. Whereas, the purely electromagnetic Transverse Magnetic (TM) surface mode  neither focusses the beam nor does it access fields at the coherence limit, these restrictions do not apply to the strongly electrostatic plasmonic modes. Electrostatic plasmonic modes are capable of focusing a charged particle beam as well as accelerating them at the field coherence or wavebreaking limit.

Besides the theoretical innovation of {\bf nonlinear and strongly electrostatic surface plasmonic modes}, plasmonic accelerators and PV/m plasmonics, in general, have become possible now due to two key recent advances in practical considerations brought about by experimental advances. 

\begin{enumerate}[topsep=4pt, itemsep=0.3ex, partopsep=0.3ex, parsep=0.3ex, label=\scriptsize$\blacksquare$]
\item {\bf Firstly}, ultrafast magnetic switching experiments observed no damage to a conductive sample after its interaction with an ultrashort, less than 100fs long nC electron beam. While no damage was detected using charge-contrast transmission electron microscopy (TEM), the free electrons responded to the beam fields and effected ultrafast magnetic switching which was observed from spin-contrast TEM. The relativistic plasmonic model offers an explanation to this observed heat deposition, beyond the Ohm's law. 

\item {\bf Secondly}, rapid and ongoing technological innovations have now made it possible to compress a charged particle beam with $\rm 10^{10}$ particles to sub-micron overall dimensions. It is modeled that a 2 nano-Coulomb electron beam can be compressed to 100nm bunch length within the next decade. Similarly, there are certain types of magnetic lens based focusing systems that are being prototyped to compress the beam to 100s of nm waist-size.
\end{enumerate}
 
In addition to these vital understructures of PetaVolts per meter plasmonics, several other theoretical, computational and experimental considerations that are crucial for its realization are identified and briefly discussed. 

The ongoing efforts to experimentally address various immediate challenges as well as to realize the short-term goals are described. A strategy is mapped out to focus on challenges to learn and incorporate principles derived from the current efforts. In addition, the longer-term challenges associated with realization of PV/m plasmonic modules for a future plasmonic collider using PetaVolts per meter gradients are outlined.

\pagebreak
\section{Grand challenge of ultra-high electromagnetic fields}
\sectionmark{Grand challenge of ultra-high gradient}

Our initiative on {\bf PetaVolts per meter plasmonics} \cite{plasmonic-3D, spie-2021} and {\bf plasmonic accelerators based future colliders}, comes at a time and in the context of reinforced commitment by the worldwide high-energy physics (HEP) community to support basic research and development (R\&D) on high gradient accelerators in recognition of their critical need for future colliders as well as fundamental discoveries, in general. 

This contribution to the proceedings of the US Community Study on the Future of Particle Physics or Snowmass 2021 process is aimed at bringing to the attention of the worldwide HEP community and the accelerator community, in particular, our initiative on a {\bf plasmonic accelerator} and its potential to open access to {\bf PetaVolts per meter} electromagnetic fields. Our work focusses on utilizing the promise of access to PV/m plasmonic fields for acceleration and focusing of particle beams.

We also highlight the strong alignment of {\bf technological trends} towards micron and sub-micron level charged particle beam compression and diagnostic techniques with the PetaVolts per meter nano-plasmonics initiative \cite{pct-2021} for future colliders and exploratory basic science and technology, in general.

Early pioneer of HEP and experimental nuclear physicist, Robert Hofstadter (awarded Nobel prize in physics in 1961) was the first to envision and advocate technological innovations to dramatically reduce the size of accelerators. It is to be noted that the experiments led by Hofstadter had extensively utilized the two-mile long Stanford linac for pioneering electron scattering studies to determine the nucleonic structure. In a proposal presented in 1968 \cite{atomic-accelerator}, Hofstadter envisioned shrinking the several mile long tunnel of his experiments to a room-sized lab using an ``atomic accelerator''. We quote \cite{atomic-accelerator}: \\ ``{\it To anyone who has carried out experiments in physics with a large modern accelerator there always comes a moment...wishes that a powerful compression...could take place.}'' and, \\
``{\it Unless there is a new idea which can radically change this picture there would seem to be some kind of limitation on the types of leviathans that the world physics can build in the future.}''

As radiofrequency (RF) acceleration technology has been a reliable workhorse of HEP since 1950s, it has undergone continual improvements and redesigns to optimize its size and economic viability. But, it has become apparent over the past several decades that this legacy technology has inherent limits on attainable energy in a practical layout for future colliders. The lack of its scalability is because the highest electromagnetic fields that RF cavities can support are limited to $\rm 100 ~ MV/m$. This limit on  attainable electromagnetic fields and the acceleration gradient dictates that the only way forward in particle physics is to increase the infrastructure size and as a consequence the cost. Moreover, this limit on gradients for both acceleration and focusing severely limit the possibilities for exploration across the Energy, Intensity and Cosmic frontiers.

Consequently, past and ongoing strategic planning processes around the world have recognized the critical need for advanced accelerator R\&D as summarized below (extracts from a few prominent reports are cited):

\begin{enumerate}[topsep=4pt, itemsep=0.3ex, partopsep=0.3ex, parsep=0.3ex]
\item {\bf 2014 Particle Physics Prioritization Planning Panel (P5) report} \cite{p5-report}: ``{\em Recommendation 26: Pursue accelerator R\&D ... with an appropriate balance among general R\&D, directed R\&D, and accelerator test facilities and among short-, medium-, and long-term efforts. Focus on outcomes and capabilities that will dramatically improve cost effectiveness for mid-term and far-term accelerators}'',
\item {\bf 2020 Update for European strategy for particle physics} \cite{eu-report}: ``{\em 3. High-priority future initiatives: A. the particle physics community should ramp up its R\&D effort focused on advanced accelerator technologies...}'',
\item {\bf 2015 HEP Advisory Panel (HEPAP) sub-panel on accelerator R\&D} \cite{hepap-report}: ``{\em Long-term: Perform the exploratory research aimed at developing new concepts that will make possible the complementary accelerator facilities of possible science interest after the Next Step}'' and ``{\em R\&D that provides higher performance at lower cost should be more heavily emphasized in the definition of the R\&D programs.}''
\end{enumerate}

{\bf Contributed by P. Muggli:} Increasing the accelerating gradient is especially critical for HEP applications that are reliant on beams with TeV energy and beyond. With gradients of the state-of-the-art RF technology being limited to $\rm <100 MeV/m$, TeV machines would be tens of kilometers long. The cost of the tunnel of length ``L'' hosting the accelerators has been estimated in a global study to be $\rm \sim2 ~ Billion ~ USD(2014) \times (L/10km)^{1/2}$ \cite{collider-cost-model}. Even with reduction in length by a factor of only a few, savings in Billions of USD are expected. A very lage reduction can be expected from the ultra-high gradients of plasmonic accelerators described here. Moreover, longer the accelerators the more complex are the associated practical constraints. For instance, stretches of land with geological characteristics suitable for the construction, operation and stability of a collider are difficult to find. Access for maintenance and repair over such distances becomes a major logistic challenge. Therefore, reducing the accelerator and collider length by increasing the accelerating gradient will not only reduce the cost of such a machine, but also simply make is possible to be built. 

\section{Scope and structure}

Plasmonic accelerator effort [1-3] uncovers an unprecedented acceleration mechanism that utilizes the unique properties of condensed matter materials with periodic ionic lattice such as certain crystals and nanomaterials to open up ultra-high gradients beyond the reach of existing techniques. Particularly, of importance are those properties that stem from the structure of the {\bf ionic lattice} which inevitably gives rise to {\bf electronic energy bands}. In conductive materials, electrons inherently occupy conduction band energy levels and are free to move in response to external excitation. Plasmonic modes are sustained by collective oscillations of the free electron Fermi gas \cite{Plasmonics-electron-gas}.

Our scope is limited to mechanisms where there exists a strongly correlated ionic lattice and corresponding electronic energy bands during these collective oscillations. Under these conditions, quantum mechanical effects dominate. Specifically, our effort focusses on acceleration and focusing gradients as high as PetaVolts per meter that can be sustained by plasmonic modes of collective oscillations of the free electron Fermi gas. 

The scope of our work {\bf DOES NOT} include:
\begin{enumerate}[topsep=4pt, itemsep=0.3ex, partopsep=0.3ex, parsep=0.3ex, label=\alph*.]

	\item {\bf Solid-state plasmas}: In solid-state plasmas the individual ions are uncorrelated as there is no ionic lattice. Solid-state plasmas are obtained by ablation of solids such as by interacting the solid with a high-intensity optical (near-infrared) laser. These plasmas have been experimentally studied in the past for laser-plasma ion acceleration in addition to being proposed for schemes such as channeling acceleration to guide and accelerate positively charged particles \cite{channeling-collider} or solid-state plasma acceleration. For further clarification we quote \cite{channeling-collider}: \\``{\it To use such... oscillations to accelerate charged particles... is problematic since the radiation length for electrons and positrons is of order 1 cm in solids,... These problems can be substantially mitigated for heavy positively charged particles by utilizing the channeling phenomenon in crystals.}'' and \\ ``{\it The basic concept of crystal channel acceleration combines plasma wave acceleration with the well known channeling phenomenon...}''
	\item {\bf Dielectric or insulating solids}: Dielectric or insulating materials have near zero electron density in the conduction band. The excitation of dielectrics relies on polarization of ionic sites which effectively generates high frequency electromagnetic radiation. It has been experimentally demonstrated that when free electrons are driven by high electromagnetic fields to transition into the conduction band of a dielectric material, the electromagnetic wave is rapidly shielded which puts an end to dielectric acceleration \cite{dielectric-damping}.
	\item {\bf Optical plasmons}: In conventional plasmonics, a low-intensity optical femtosecond laser is used to excite weakly-driven plasmons. Such weakly-driven optical plasmons have also been proposed as a novel pathway for particle acceleration \cite{optical-plasmon}. On the other hand, the longer timescale radiation energy that is unavoidably present in the pre-pulse of a high-intensity optical laser ablates the interacting material into solid-state plasma which is out of the scope of our work.

\end{enumerate}
While the existing parallel efforts on R\&D of plasma and dielectric accelerators have been duly recognized and well supported by funding agencies, our effort being at an early stage calls out for special attention and investment by the HEP community.

In order to bring out the promise of this effort, while also identifying its challenges and mapping out strategies for the future, the following sections are structured as follows. 
\vspace{-1.0mm}
\begin{itemize}[topsep=4pt, itemsep=0.3ex, partopsep=0.3ex, parsep=0.3ex, label=\scriptsize$\blacksquare$]
\item Sec.\ref{non-collider-HEP} charts out a physics case for the possibility of using the PetaVolts per meter fields of Plasmonic modes over the next decade to open up exploration at the frontiers of high-energy physics. 
\item Sec.\ref{plasmonic-acc-fundamental} introduces the fundamentals of plasmonic accelerators and the ultrafast dynamics of free electron Fermi gas.
\item Sec.\ref{ultrafast-expt-conducting-materials} ({\em contributed by J. Stohr}) summarizes the certain fascinating findings of a past experiment that specifically studied the interaction of an ultrafast electron beam with conducting materials.
\item Sec.\ref{surface-crunch-in-plasmonic} provides brief description of the plasmonic modes that enable acceleration and focusing of particle beams. The role of a strongly electrostatic surface plasmonic mode to mitigate well known disruptive effects of collision between particle beams and bulk materials is also briefly discussed.
\item Sec.\ref{extreme-bunch-compression} ({\em contributed by G. White and G. Andonian}), discusses novel techniques for extreme bunch compression, focusing and diagnostic for ultra-dense particle beams. This includes longitudinal compression towards $< 100 \rm nm$ bunch lengths and multi-MegaAmpere peak currents as well as focusing of the beam waists to the order of $100 \rm nm$.
\item Sec.\ref{xray-laser-nanostructures} ({\em contributed by G. Mourou and T. Tajima}) describes x-ray laser driven nanostructures which use the proposed high intensity keV-photon laser to drive carbon nanotubes.
\item Sec.\ref{challenges-prototype}, describes the challenges, possibilities and efforts (ongoing and future) to prototype plasmonic accelerators.
\end{itemize}

\section{Non-collider searches through direct access of HEP frontiers: \\ Physics case}
\label{non-collider-HEP}
\sectionmark{Physics case for non-collider tests of HEP frontiers}

In this section we chart out short-term physics case for PetaVolts per meter plasmonics initiative. This is based upon non-collider examination of the predictions of theoretical models in HEP that lie at the energy, intensity and cosmic frontier. With conventional and other advanced mechanisms of producing energetic particles, these non-collider searches remain out of the reach. However, the availability of PV/m plasmonic fields will for the first time make it possible to pursue these predictions. 

These are the {\bf decadal goals of the PetaVolts per meter plasmonics} effort in parallel with understanding various elements of an integrated collider design using plasmonic accelerator modules.

\subsection{Examination of nonlinear Quantum Electro-dynamics (QED): \\ opening the vacuum with extreme plasmonic fields}

It is predicted that extremely strong electromagnetic fields produce positron-electron pairs directly off of the vacuum when their energy density is high enough. Specifically, the QED theory that when the electromagnetic fields approach the Schwinger field limit of $\rm E_s = 1.3 \times 10^{18} V/m$ \cite{schwinger-limit}, the probability of observing positron-electron pairs directly produced from the vacuum increases with the electric field, $\rm E_{lab}$ as $\rm exp(-\frac{\pi m_e^2}{eE_{lab}})$, as is the case with tunneling effects. 

PetaVolts per meter plasmonics can ``open the vacuum'' based upon plasmonic nano-focusing of the beam \cite{plasmonic-nanofocusing}. While there are existing lab-based routes to experimentally examine this fundamental HEP theory of spontaneous pair-production, such as colliding two high-intensity laser pulses or the collision of an ultra-relativistic particle beam with a laser pulse, PV/m plasmonics offers certain distinct experimental capabilities.

As the plasmonic modes have a near-zero group velocity and are localized in space, the plasmonic fields {\bf DO NOT}:
\begin{itemize}[topsep=4pt, itemsep=0.3ex, partopsep=0.3ex, parsep=0.3ex]
 \item travel with the particle or laser beams, or 
 \item are experienced in a Lorentz-boosted frame, or
 \item exist only momentarily at the collision
\end{itemize} 
unlike the underlying construct in other experimental techniques. Moreover, it is easier to tune the interaction parameters of the plasmonic experiments. 

\subsection{Examining quantum gravity model: \\ ultra-energetic photons from nano-wiggler}

Theories that attempt to describe gravity using a quantum mechanical framework as well as theory of everything models are considered to be examinable only using astrophysical observation data. One of predictions of the string theoretical model is that the vacuum exhibits a non-trivial permittivity to ultra-energetic photons \cite{quantum-gravity}. With the nature of the interaction between short wavelength photons and the vacuum, the most appropriate probes for this model are sources that produce short bursts of ultra-energetic photons.

However, with the existing state of technology the only bursty sources of TeV or PeV photons are astrophysical bodies. Therefore, detection of these quantum gravity effects is sorely dependent on observational data. 

Production of TeV photons using the nano-wiggler mechanism \cite{plasmonic-3D} based upon PV/m plasmonics promises to open novel possibilities for searches in HEP. The nano-wiggler utilizes the ultra-high focusing fields of plasmonic modes to wiggle the ultrarelativistic electrons of the beam that have a relativistic factor, $\rm\gamma_b\simeq 10^{4-6}$ with nanometric wiggling oscillations wavelengths, $\lambda_{\rm osc}$. The energy of the photons so produced is $\rm 2\gamma_b^2 ~ hc/\lambda_{\rm osc}$.

\section{PetaVolts per meter Plasmonics fundamentals}
\label{plasmonic-acc-fundamental}

Plasmonic accelerators \cite{plasmonic-3D, spie-2021} rely on the electronic energy band structure inherent in conducting materials due to an appropriate combination of suitable constituent atoms and ionic lattice structure. The electron energy band structure of conducting materials supports free electrons (electron not tied to any specific atoms) in the conduction band. In particular, electrons that occupy energies in the conduction band constitute a {\em free electron Fermi gas}.

The Fermi electron gas freely moves about the entire lattice. It is interesting to note that the free electron Fermi gas in conducting solids is the highest density collection of electrons that is accessible at a terrestrial lab. Upon external excitation, this electron gas undergoes collective oscillations \cite{Plasmonics-electron-gas} and in the process excites collective fields which form the basis of the general field of plasmonics and plasmonic accelerators, in particular. 

\begin{figure}[!htb]
\centering
   \includegraphics[width=0.5\columnwidth]{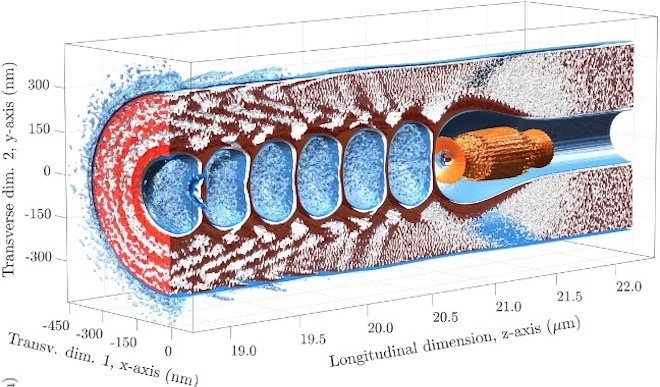}
   \caption{Density profile of free electron Fermi gas of surface crunch-in plasmonic mode, reproduced from \cite{plasmonic-3D}, in a tube with equilibrium conduction band density, $n_e$ in the tube walls of $n_t = \rm 2 \times 10^{22}cm^{-3}$. The snapshot is at $\mathrm{20\mu m}$ ($\sim$ 73fs) of interaction of a $\sigma_z\mathrm{=400nm}$, $\sigma_r\mathrm{=250nm}$ beam with tube. The tube has a vacuum-like core of radius, $r_t\mathrm{=100nm}$ (nearly flat-top beam limit, $\sigma_r= 2.5\times r_t$). The beam envelope (in orange) although initially larger than the tube has undergone self-focusing approaching ultra-solid densities.}
\label{fig:3D-crunchin-mode-beam-tube}
\end{figure}

The strongly electrostatic surface plasmonic mode underlying a plasmonic accelerator which is sustained by relativistically oscillating Fermi electron gas is depicted in Fig.\ref{fig:3D-crunchin-mode-beam-tube} reproduced from \cite{plasmonic-3D}. This figure is a snapshot in time of the evolving interaction between an electron beam (with its envelope at a given density in orange) and the free electron Fermi gas (darker colors representing a higher density) modeled using simulations that use the particle tracking approach to treat collisionless behavior of the conduction band free electron gas.

The free electron Fermi gas is a quantum mechanical entity. The near-continuum energy levels that make up the conduction band and give rise to free electrons are entirely a consequence of two distinct quantum mechanical phenomena related to the electron wave-function in an ionic lattice:
\begin{enumerate}[topsep=4pt, itemsep=0.3ex, partopsep=0.3ex, parsep=0.3ex, label=\roman*.]
\item  inter-atomic bonding due to orbital overlap of neighboring atoms, and 
\item periodic potential exerted by the background lattice on electron wave-functions. 
\end{enumerate}
Due to their quantum mechanical origins and having a fixed energy (and dispersion relation), collective oscillations of free electron Fermi gas are treated as quasi-particles and labelled as plasmons.

Plasmonic accelerators rely on the excitation of the free electron Fermi gas by ultrafast sources with temporal duration that may perturb or distort the ionic lattice but cannot disrupt it. Modeling the transient modification of the ionic lattice and as consequence the electron energy band structure has been identified as a key challenge which is summarized in sec.\ref{challenges-prototype}. 

Over those timescales of interaction between a charged particle beam and the free electron Feri gas, where the periodic ionic lattice structure and its corresponding quantum mechanical derivative of electronic energy band structure are still present, plasmonic processes are dominant and control the interaction. Especially because the free electron Fermi gas has the highest density and therefore (in accordance with Eq.\ref{eq:plasmonic-oscillation-time}) the shortest response times to external excitations. Therefore, plasmonic processes are the fastest when considering electron dynamics.

\vspace{5.0mm}
\noindent{\bf Properties of plasmonic modes:} As a consequence of the fact that the equilibrium density of free electron Fermi gas in conducting solids defines the limit of accessible electron density, its dynamics presents a unique set of properties. 

The characteristic energy of a plasmonic quasi-particles depends only on the free electron density $n_e$ in the conduction band of an ionic lattice,  $\hbar\omega_{\rm plasmon}$ ($=3.7{\rm eV} ~ \sqrt{n_e \rm [10^{22} cm^{-3}]}$) . For comparison, the photon energy of a near-IR optical ($\lambda_0=800\rm nm$) laser is around 1.2eV.

The characteristic oscillation time of the free electron Fermi gas is,
\begin{align}\label{eq:plasmonic-oscillation-time}
\begin{split}
	2\pi \omega_{\rm plasmon}^{-1} [{\rm fs}] = 1.1 ~ \left[n_e \rm (10^{22} cm^{-3}) \right]^{-1/2}
\end{split}
\end{align}

The characteristic spatial dimension or size of plasmons is dictated by their wavelength which depends upon the properties of the conducting condensed matter material that sustain them. The wavelength of plasmons is a function of the density ($n_e$) of the free electron Fermi gas in the conduction band, 
\begin{align}\label{eq:plasmonic-wavelength}
\begin{split}
	\lambda_{\rm plasmon} {\rm [nm]} = 330 ~ \left[n_e \rm (10^{22} cm^{-3}) \right]^{-1/2}
\end{split}
\end{align} 
In a typical metallic nanostructure with conduction band electron density of $\rm 10^{22-24} \rm cm^{-3}$, the characteristic size of a plasmon (planar surface plasmon is longer) ranges from few hundred nanometers to a few nanometers.

If the free electron Fermi gas is undergoing relativistic oscillations with electron momentum of $\vec{p}_{\rm plasmon}$ (and velocity, $v_{\rm plasmon}$) approaching $m_ec$ as is the case in our work, then the time and spatial scales of plasmonic oscillations need to be modified (and elongated) by the average relativistic factor of electrons, $\gamma_{\rm plasmon}=\sqrt{1 + \frac{\vec{p}}{m_ec}^2 }=(1-\beta^2)^{-1/2}$, where $\beta=v_{\rm plasmon}/c$, as follows:
\begin{align}\label{eq:relativistic-plasmonics}
\begin{split}
	2\pi \omega_{\rm plasmon}(\gamma_{\rm plasmon})^{-1} [{\rm fs}] = 1.1 ~ \sqrt{\gamma_{\rm plasmon}} ~ \left[n_e \rm (10^{22} cm^{-3}) \right]^{-1/2} \\
	\lambda_{\rm plasmon}(\gamma_{\rm plasmon}) {\rm [nm]} = 330 ~  \sqrt{\gamma_{\rm plasmon}} ~ \left[n_e \rm (10^{22} cm^{-3}) \right]^{-1/2}
\end{split}
\end{align} 

Highly non-linear electron oscillations of the electron gas approach the coherence limit of orderly collective electron oscillations or the wave-breaking limit \cite{nonlinear-oscillations} which is defined in terms of the electric field amplitude that is sustained during oscillations. This coherence limit of the electric field is,
\begin{align}\label{eq:plasmonic-fields}
\begin{split}
	{\rm E}_{\rm plasmon} \rm [TV/m] = 9.6 ~ \sqrt{n_e (10^{22} \rm cm^{-3})}. 
\end{split}
\end{align}
While electric fields a few times the coherence limit are achievable, the electron oscillations become less orderly and coherent upon exceeding the limit in Eq.\ref{eq:plasmonic-fields}.

\vspace{5.0mm}
\noindent{\bf Key enablers behind the emergence of plasmonic accelerators:}
There are two fundamental realizations that have enabled the emergence of ultra-high gradient plasmonic accelerator modules for a future collider:
\begin{enumerate}[topsep=4pt, itemsep=0.3ex, partopsep=0.3ex, parsep=0.3ex, label=\roman*.]
\item past experimental observation of the {\bf absence of damage in conducting materials} for electron bunch lengths, $\sigma_{\parallel} \leq 10\mu m$ as summarized in section \ref{ultrafast-expt-conducting-materials}, and 
\item {\bf sub-micron bunch compression techniques} (discussed in sec.\ref{extreme-bunch-compression}) that can match the spatial, Eq.\ref{eq:plasmonic-wavelength} and temporal, Eq.\ref{eq:plasmonic-oscillation-time} scales of plasmonic oscillations: (i) which can longitudinally compress the bunch through manipulation of the bunch phase-space correlations using magnetic lattices, (ii) bunch waist compression using permanent magnet quadrupole triplet systems.
\end{enumerate}

\vspace{5.0mm}
In contrast with conducting media which have free electrons occupying the conduction band, insulators or dielectrics have near zero electron density in the conduction band. So, plasmonic modes can not be sustained in insulating materials.
 
\noindent{\bf Distinction from Dielectric or Insulating solids:} ({\it contributed by T. Katsouleas})
Although there appear to be phenomenological similarities between the plasmonic modes in conducting media and the dielectric modes in an insulting dielectric because both are excited in a hollow solid tube, the physics of each is quite distinct.  

Whereas plasmonic modes are supported by collective oscillations of free electron Fermi gas, dielectric modes are supported by polarization currents generated by the distortion of the electron cloud of ions in the lattice. Plasmonic oscillations are out of phase with the drive electric field while dielectric polarization currents are not. In dielectrics, polarization currents excite large amplitude Cherenkov radiation which supports the acceleration mechanism. In plasmonics, electrostatic plasmonic modes supported by large-scale charge separation between the free electron Fermi gas and ionic lattice sustain ultra-high gradients.

As a result the dispersion relation of each type of wave is quite different:  
\begin{itemize} 
	\item $\omega = \omega_{\rm plasmon}/\sqrt{2}$ for a plasmonic surface wave and 
	\item $\omega = k~c/n_{\rm dielec}$ for a dielectric wave, where $n_{\rm dielec}$ is the index of refraction of the dielectric.
\end{itemize}

\begin{figure}[!htb]
\centering
   \includegraphics[width=0.5\columnwidth]{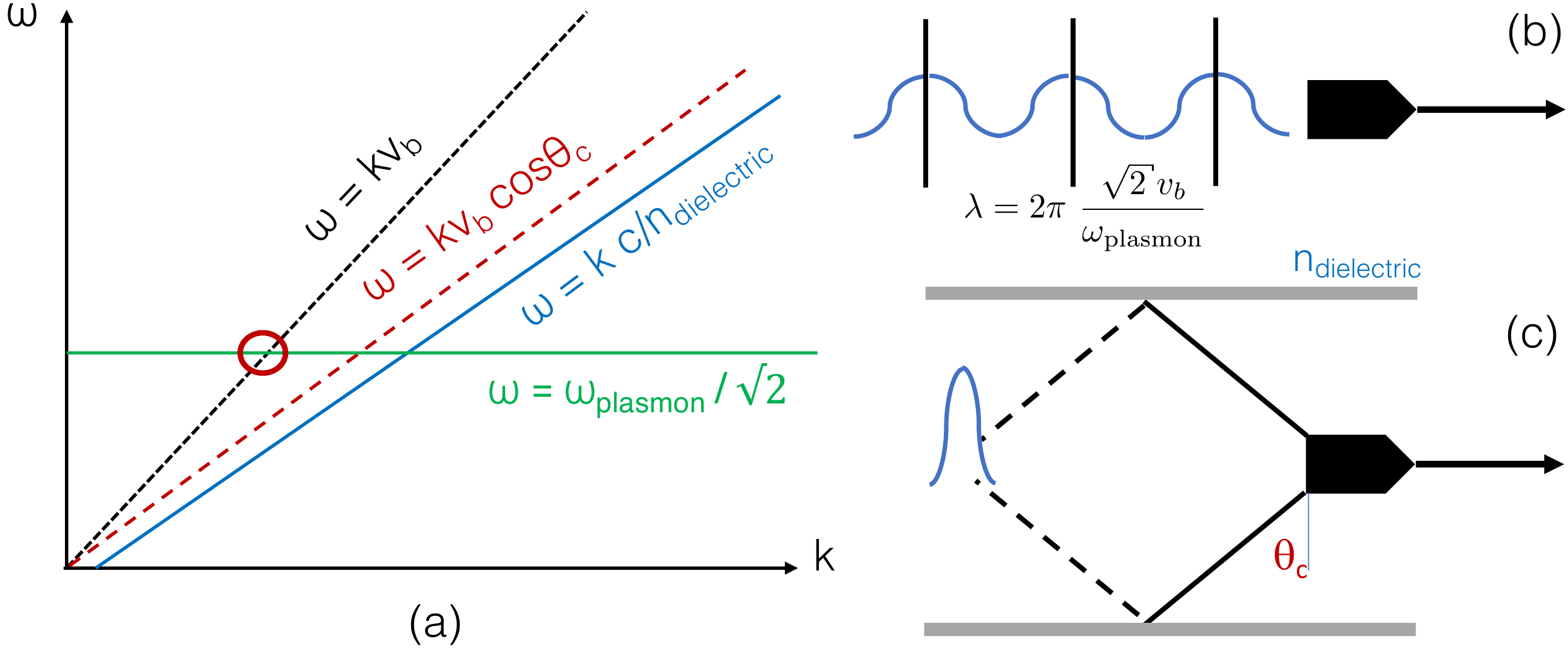}
   \caption{Dispersion relation for surface plasmon and dielectric modes in (a) and their overlap with the drive particle bunch which determines the spatial and temporal profile, (b) of the surface plasmon mode, (c) of the dielectric mode. }
\label{fig:dispersion-plasmonic-dielectric}
\end{figure}

The resulting mode structure can be predicted from the intersection of the dispersion relations and the driver disturbance at $\omega = k v_b$ (where, $v_b$ is the beam velocity) as in Fig.\ref{fig:dispersion-plasmonic-dielectric}a.  From this we see that the surface plasmonic mode has but one wavenumber and frequency and gives rise to planar wave fronts and sinusoidal oscillations as seen in the cartoon of Fig.\ref{fig:dispersion-plasmonic-dielectric}b.  On the other hand, the dielectric dispersion intersects the beam disturbance at the Cerenkov angle ($cos(\theta_c) = 1/n_{\rm dielec}$) at every $\omega$ and $k$.  This supports the familiar Cerenkov cone type wavefront structure familiar in the wake of supersonic jets and illustrated in the cartoon in Fig.\ref{fig:dispersion-plasmonic-dielectric}c.  

This shock-like cone propagates out to the outer conducting wall boundary of the dielectric whereupon it is reflected back toward the axis, giving rise to an accelerating spike on the axis as in Fig.\ref{fig:dispersion-plasmonic-dielectric}c.  It is clear from this picture that these two modes differ considerably. In the first, the accelerating field is nominally sinusoidal (in the linear regime) and the location of the accelerating peak is controlled by the free electron Fermi gas density in the tube. In the second, the accelerating field is a single spike and its location is determined by the outer radius of the dielectric and the dielectric constant of the tube.  

\noindent{\bf Distinction from Plasmas:}
It is quite critical to note that our initiative does NOT deal with solid-state plasmas or gaseous plasmas. In solid-state plasmas, produced by ablation of solids the ionic lattice is completely disrupted with the constituent ions entirely losing their mutual correlations. By definition, solid-state plasmas therefore, do not have an energy band structure. Solid-state plasmas have been utilized for laser-plasma ion acceleration, but, due to their contingency on the turbulent ablation process they may not be controllable for consistent properties desired from acceleration modules underlying a collider. 

Due to highly randomized nature of ions in gaseous plasmas, the properties of plasma oscillations significantly diverge from that of plasmonic modes especially at solid densities. Using one simplified example, the characteristics of electron-electron collision with increase in the density of a gaseous plasmas are ineffectual for a {\bf quantum mechanical electron gas} due to strict requirements set by the {\bf Pauli exclusion principle}.

\vspace{5mm}
\section{Past ultrafast experiments with conductive materials \\ {\normalsize \it contributed by J. Stohr} }
\label{ultrafast-expt-conducting-materials}
\sectionmark{Past ultrafast experiments with conducting materials}

A set of past experiments \cite{Stohr-FFTB-2003,Stohr-FFTB-2009} that studied the interaction of conducting materials (Iron/ Cobalt alloys) with ultrafast electron beam of Final Focus Test Beam \cite{FFTB-compression} at the Stanford linear accelerator center (SLAC) sets the precedent to answering several questions about plasmonic accelerator in addition to beam-driven plasmonics, in general. 

The findings of these experiments are directly relevant and provide a strong experimental basis for the initiative on plasmonic accelerators. This is because these past experiments on ultrafast magnetic switching are also based upon femtosecond excitation \cite{Stohr-Siegmann-book} of metallic alloys that possess free electron Fermi gas.

Experiments conducted on ultrafast magnetic switching using the FFTB electron bunch at the SLAC laboratory observe fascinating phenomenon in a metal, namely {\bf the lack of heating}, when the field strengths are increased well into the GV/m regime while the field pulses are shortened to less than 100 fs. These experiments studied the response of the same $\rm Co_{70}Fe_{30}$ thin film samples of dimensions and geometry described in \cite{Stohr-FFTB-2003}. 

\begin{figure}[!htb]
\centering
   \includegraphics[width=0.37\columnwidth]{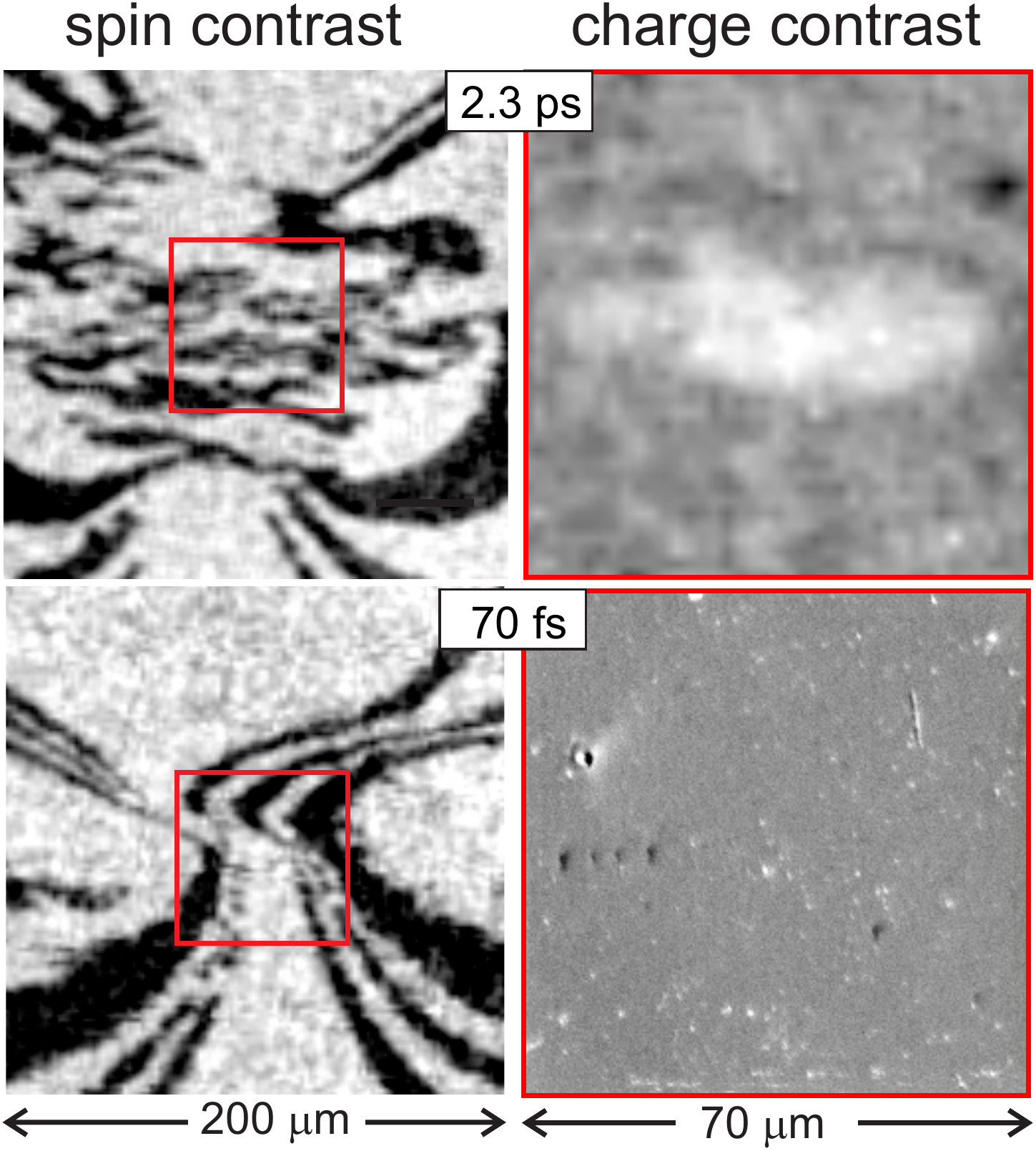}
   \caption{A comparison of the sample images obtained with Scanning Electron Microscope (SEM) with spin polarization (SEMPA), on the left, and conventional SEM pictures of the structure, on the right, zooming in on the region where the beam hit. It is evident from this comparison that for the shorter electron pulses, $\sigma_{\parallel}\leq \rm 100fs$ no topological deformation is seen.}
\label{fig:Stohr-beam-damage}
\end{figure}

Fig.\ref{fig:Stohr-beam-damage} a comparison of material damage for the two field pulses of different length and strength illustrated using spin sensitive transmission electron microscopy in Fig.\ref{fig:Stohr-beam-damage}(a) and conventional transmission electron microscopy in Fig.\ref{fig:Stohr-beam-damage}(b).

By examination of the magnetic patterns written into the ferromagnetic film by the fields of the electron bunch we find that only the longer electric fields of 2.3 ps and magnitude $\rm\leq 0.6 GV/m$ heat the sample above its Curie temperature of $T_C \leq 1200K$, while no such heating is observed for the shorter 70fs pulse with fields up to 20GV/m.

As in previous work the experiments were carried out using the Stanford Linear Accelerator and by use of a special electron bunch compressor the Gaussian shaped bunch length could be shortened from several picoseconds to below 100 femtoseconds. We used bunches with longitudinal Gaussian rms widths, $\sigma_t$ of 2.3 ps and 70 fs, as indicated in Fig.\ref{fig:Stohr-beam-damage}. As a result of relativistic field contraction, the fields are entirely parallel to the surface of the sample oriented perpendicular to the direction of bunch propagation. Our 10nm thick polycrystalline $\rm Co_{70}Fe_{30}$ thin film sample with uniaxial in-plane magnetic anisotropy was deposited onto a 0.5 mm thick [110]-MgO substrate with intervening buffer layers consisting mainly of 30nm $\rm Cr_{80}Mg_{20}$ and capped with a 1.5 nm thick layer of Pt to prevent corrosion.

The magnetic pattern written by the 2.3 ps pulse exhibits the formation of stripe domains along the horizontal easy axis which are characteristic for a sample that has been heated to a temperature above the Curie temperature $T_C = 1200 K$ and then cooled in the absence of a magnetic field. In addition, conventional SEM images revealed non-magnetic changes of the sample surface near the beam impact area, as shown on the right in Fig. \ref{fig:Stohr-beam-damage}. In contrast, the magnetic pattern written by the 70 fs pulse, shown underneath, has sharp domain walls and is clearly written by the magnetic beam field as discussed in the earlier results with 2.3 ps pulse \cite{Stohr-FFTB-2003}. Also, no non-magnetic SEM contrast indicative of beam ``damage'' was observable.

The pulsed E-field leads to current flow and Joule heating. Since the peak E-field of the beam scales as $N_e/\sigma_t$ and the pulse energy scales with $N_e^2/\sigma_t$ (where, $N_e$ is number of bunch particles and $\sigma_t$ is the bunch length), the shorter pulse is expected to produce more heat. This is at odds with our experimental observation. 

It is surmised that these puzzling observations can be explained by the non-linear response of the electrons in a metal when the fields become extremely large. The relativistic and nonlinear response of the free electron Fermi gas is driven by GV/m electric fields of the beam results in processes that do not follow the conventional Ohm's law.

\section{Strongly electrostatic and relativistically oscillating plasmons}
\label{surface-crunch-in-plasmonic}
\sectionmark{Strongly electrostatic and relativistic plasmons}

\noindent{\bf Surface plasmon:} For preservation of the essential properties of a particle beam, it is critical to mitigate its direct collision with the ionic lattice. Random collisions between particles and the ionic lattice not only result in loss of energy of the individual particles of the beams but can also disrupt the beam, as a collective entity, through a wide-range of instabilities. These instabilities include filamentation, hosing and several others which can rapidly grow and completely disrupt the beam. Our effort on plasmonic accelerators, therefore, utilizes a surface plasmon where the beam propagates through vacuum surrounded by conducting walls.

\vspace{5mm}
\noindent{\bf Conventional surface modes:} It is well known that the Transverse Magnetic (TM) mode which is a purely electromagnetic mode sustains zero focusing forces on a beam inside a metallic cavity or tube. In fact contrary to focusing forces, charged particle beams propagating inside hollow cavities are well known to excite strong deflecting forces if not perfectly aligned to the axis of the cavity.

Whereas the conventional metallic cavities that form the basis of modern particle particle accelerators and light sources utilize electron currents in the metallic walls that oscillate only at the radiofrequency (rf), plasmonic accelerators utilize collective oscillations at spatial and temporal frequencies, Eq.\ref{eq:plasmonic-oscillation-time} and \ref{eq:plasmonic-wavelength}, dictated by the density of the free electron Fermi gas itself.

\vspace{5mm}
\noindent {\bf Purely electromagnetic surface wave mode:} Besides metallic cavities, existing works on excitation of the TM mode by a charged particle beam in a hollow tube of gaseous plasma have also analytically demonstrated that there are zero focusing forces \cite{Katsouleas-PRL-1998}, we quote: 

\begin{center} 
{\em ``The focusing force is zero inside the channel for a very relativistic particle."}
\end{center}

\noindent {\bf Experiments on purely electromagnetic surface wave mode:} Experiments on excitation of TM mode by a charged particle beam in a hollow tube of gaseous plasma have confirmed the above theoretical result \cite{Positron-hollow-channel-2016}, we quote:

\begin{center} 
 {\em ``We find that when the positron beam propagates on-axis through the channel, there are no significant changes in the spatial profile, indicating the absence of the transverse focusing forces within the channel, thus demonstrating the merit of using hollow channel plasmas for PWFA.''}
 \end{center}
 
\noindent The crunch-in regime \cite{crunch-in-regime,crunch-in-regime-ipac} is known to violate the conditions imposed by the purely electromagnetic TM mode.

\vspace{5mm}
\noindent{\bf Strongly electrostatic surface crunch-in mode:}
The strongly electrostatic surface crunch-in plasmonic mode, in contrast with the TM mode, sustains strong focusing forces. As the free electron Fermi gas is driven by beam fields larger than tens of MV/m, the free electron gas gains relativistic momentum. With relativistic momentum, the kinetic energy of plasmonic oscillations increases beyond the surface potential. The oscillating electron gas can thus breach the surface potential. Further details of relativistic plasmonic oscillations and the surface crunch-in process are in \cite{plasmonic-3D, spie-2021}.

\begin{figure}[!htb]
\centering
   \includegraphics[width=0.35\columnwidth]{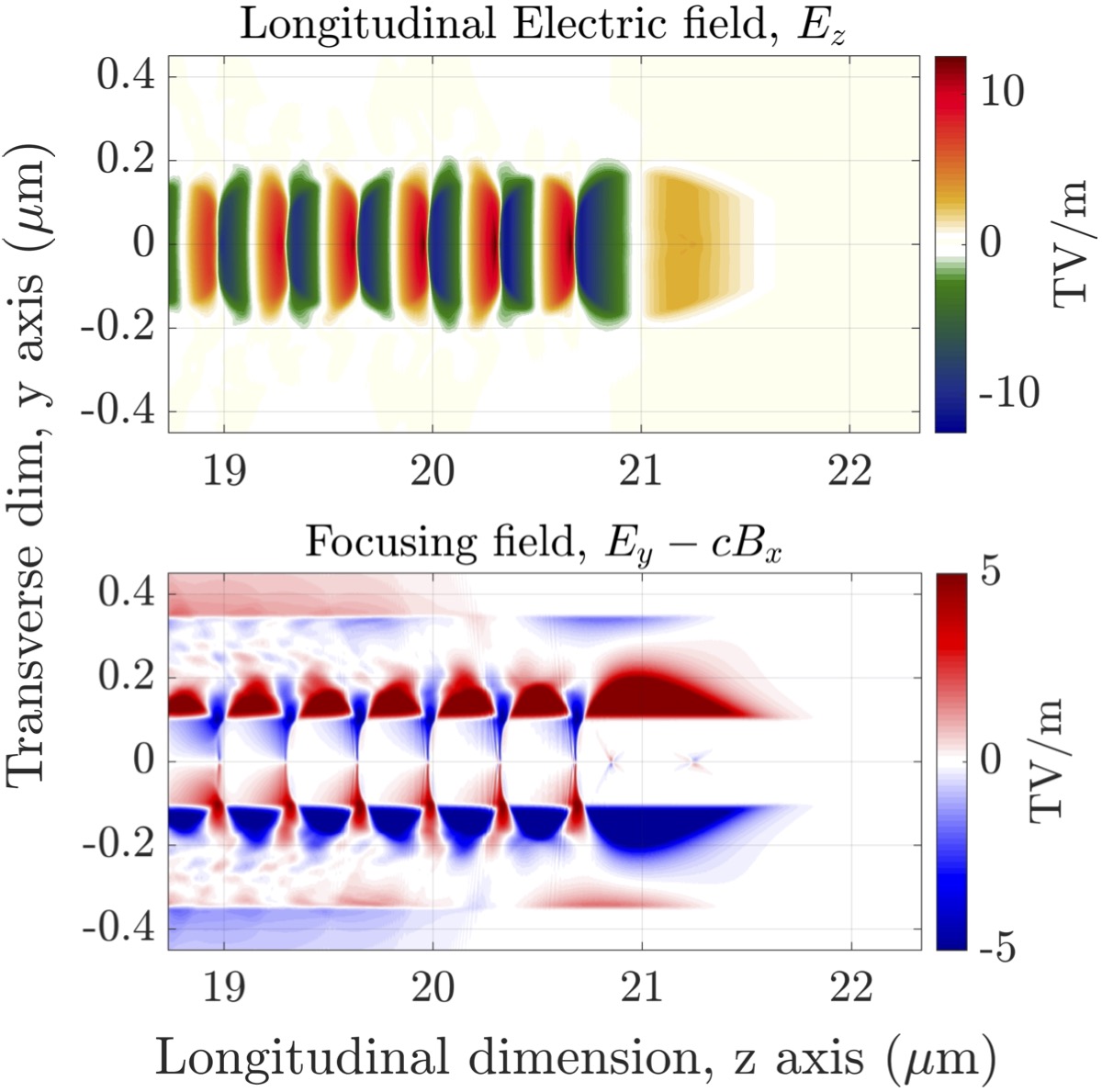}
   \caption{Longitudinal (top, $\rm E_z$) and Focusing (bottom, $\rm E_y-cB_z$) fields of the surface crunch-in plasmonic mode from 3D PIC simulations in cartesian geometry. The tube and beam parameters are exactly the same as in Fig.\ref{fig:3D-crunchin-mode-beam-tube}.}
\label{fig:3D-crunchin-fields-beam-tube}
\end{figure}

The density structure of the free electron Fermi gas oscillating to sustain the surface crunch-in mode is demonstrated in Fig.\ref{fig:3D-crunchin-mode-beam-tube} and the corresponding field structure is shown in Fig.\ref{fig:3D-crunchin-fields-beam-tube}. Due to the highly nonlinear and relativistic nature of plasmonic oscillations in our work, crunch-in plasmonic mode is demonstrated using particle-in-cell simulations of metallic free electron Fermi gas density of $n_t = \rm 2\times 10^{22} cm^{-3}$. The beam density is $n_b = \rm 5 \times 10^{21} cm^{-3}$. These simulations demonstrate that the surface crunch-in plasmon supports strong focusing fields ($\rm E_r$) in addition to longitudinal field ($\rm E_z$). 

As these plasmonic oscillations are sustained at the coherence limits which are dictated by the density of the free electron Fermi gas, $n_e$, they support electromagnetic fields at the coherence limit in Eq.\ref{eq:plasmonic-fields}. This is modeled to be possible by exciting novel high amplitude modes of plasmonic oscillation with dense particle beams. It is important note that in the crunch-in mode, just like the beam which excites the plasmonic mode, the accelerated bunch does not interact with the background ionic lattice. Due to this negatively charged as well as positively charged particles can be accelerated with equivalent efficacy. Apart from the injection of an electron within the plasmonic mode, it is also possible to inject ultrashort positron bunch as well as positive and negative muons. Ultrashort bunches of exotic particles, such as positrons \cite{Sahai-positron} and muons \cite{Sahai-muon}, may be obtained from a plasma-based accelerator.

\vspace{5mm}
\noindent{\bf Surface crunch-in plasmonic mode breaks the scaling of purely EM surface modes:} {\bf (contributed by T. Katsouleas)}
It follows from the Panofsky-Wentzel theorem at a given frequency that when the longitudinal field of a charge in a structure increases with miniaturization as $1/a^2$, where $a$ is the transverse structure dimension, the transverse field amplitude increases as $1/a^3$.  These transverse fields lead to head-tail instabilities that severely limit the amount of charge that can be accelerated in miniaturized devices. Moreover, transverse misalignment between the beam axis and the axis of symmetry of the structure excites higher-order mode that further deflect the beam off-axis.

On the other hand, this scaling can be overcome when the transverse aperture is dynamic as in a plasmonic crunch-in surface mode. In that case the head of an electron beam strongly drives out the free electron Fermi gas which as part of their oscillation trajectory go across the surface and re-converge on the axis behind the driver.  The aperture behind the beams can be vanishingly small and result in large accelerating fields, while the aperture seen by the beam and associated with the transverse fields is much larger.  These dynamic structures can overcome the scaling limitation described above that otherwise limits fixed structures.   

The study of the above regime of plasmonic mode in micro to nano-scale tubes will reveal the new physics of plasmonic surface modes.  Plasmonic modes potentially offer extreme gradients and an alternate and potentially more direct path to achieving milestones such as staging, positron acceleration and control of emittance growth.

\section{Extreme bunch compression, focusing and \\ diagnostics for plasmonics \\ {\normalsize \it contributed by G. White and G. Andonian}}
\label{extreme-bunch-compression}
\sectionmark{Extreme bunch compression, focusing \& diagnostics}

\subsection{Extreme compression of particle beams to greater than 100kA peak bunch current \\ {\it contributed by G. White}}
Recent advances in bunch compression has opened up the possibility of producing sub-micron bunch lengths with MegaAmpere (MA) peak currents. For example, ongoing upgrades at the FACET facility of the Stanford Linear Accelerator Center (CLAC) are expected to provide the possibility for $\rm > 200kA$ pulses with $\rm <1 \mu m$ rms bunch length in the near future. Precedented to the attainment of sub-micron electron bunches was set by routine generation of 20-30 ?m rms bunch length, with 30kA peak current at the same facility during the last decade.

Taking the next logical step and compressing electron bunches into the $\rm <0.1 \mu m$, $\rm >1MA$ regime would support the development of revolutionary new applications across a range of fields including PetaVolts per meter plasmonics and plasmonic accelerators for addressing challenges across the energy, intensity and cosmic frontiers. 

To reach the regime of multi-MA peak current compression, in addition to a next-generation high-brightness source or  injector (not discussed here), various non-linearities present in the bunch compression process need to be considered in order to preserve both longitudinal and transverse emittance. 

One of the most limiting physical processes is that of coherent synchrotron radiation (CSR) which causes transverse emittance growth of many orders of magnitude, in addition to strongly limiting the final achievable peak current in the regime considered. To compensate for the CSR emittance degradation, chicane designs more complex than the standard 4-bend chicane need to be considered, e.g. multi-bend chicanes, quadrupole and sextupole loaded chicanes or arcs or wigglers.

\begin{figure}[!htb]
\centering
   \includegraphics[width=0.5\columnwidth]{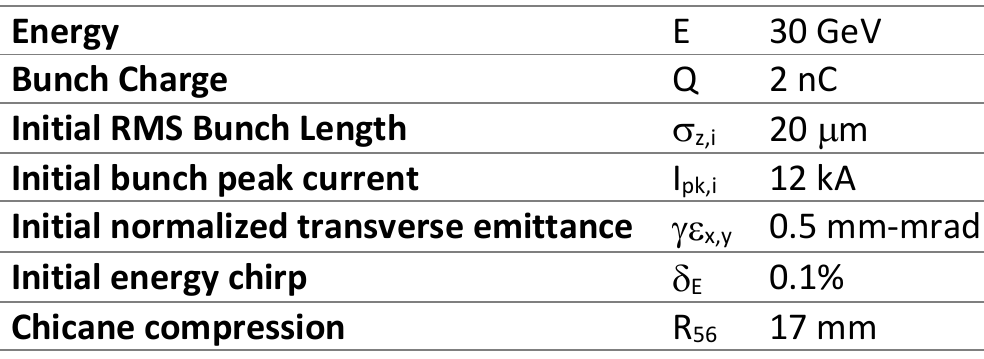}
   \caption{Initial electron bunch parameters considered for final bunch compression stage.}
\label{fig:30GeV-design-parameters}
\end{figure}

\vspace{1.5mm}
\noindent {\bf Design Parameters, Constraints and General Considerations}
\begin{itemize}
\item We consider a bunch compressor design capable of compressing an electron bunch to peak currents $\rm \gg100kA$ with $\rm \ll 1 mm-mrad$ transverse emittance growth.
\item A next-generation photo-injector with transverse emittance $<0.1 \rm mm-mrad$, 0.1-2 nC charge and few 100A initial peak current is considered. Typical injector energies are $\sim 100 \rm MeV$.
\item The injector should be capable of $\rm > 2nC$ initial charge to allow for collimation of tails as part of compression scheme.
\item Although precise amount of charge is dependent on the scientific case, most demanding is a collider which requires $\rm >nC$ bunch charge and the smallest transverse emittances, $<0.05 \rm mm-mrad$
\item	Multi-stage compression will be required to achieve compression ratios $\rm >1000\times$
	\begin{itemize}
			\item E.g., FACET-II uses 3 compression stages at 0.3, 4.5, 10 GeV
			\item Removes need for excessive energy spread
	\end{itemize}
\item Final compression stage should operate at the final energy
	\begin{itemize}
		\item Helps with CSR, where emittance growth scales as $\Delta \epsilon_n/\epsilon_n \sim 1/\sqrt{\gamma}$
		\item Adverse surface effects are expected at high compression ($\rm I^2R$ pulse heating)
		\item This allows the use of the energy chirp induced by wakefields in the final accelerating section:
			\begin{itemize}
				\item Demands final compression stage should have +ve R56 (requires quadrupole loaded optics) to make use of wakefield chirp
				\item This also helps with non-linearities from compression system, T566 due to partial self-cancelation from rf curvature
			\end{itemize}
	\end{itemize}
\item As a good compromise between CSR mitigation, whilst maintaining a reasonable system length (which scales strongly with energy due to Incoherent Synchrotron Radiation growth in bends) here a final energy of 30 GeV is considered with beam parameters tabulated in Fig.\ref{fig:30GeV-design-parameters}.
\end{itemize}

\noindent {\bf NB:} Convention used here is {\bf -ve R56} requires {\bf higher energy at tail} of bunch, {\bf +ve R56} requires {\bf higher energy at head} of bunch.

Initial compression stages reduce the initial bunch length to $\rm 20\mu m$ rms, where the peak current is a 12 kA, similar to the largest values used at FEL facilities, where at low charge emittance preservation has been experimentally demonstrated ($\rm < 200 pC$ charges). The final compression stage required to compress the electron bunch to $\rm <1 \mu m$ in length requires careful design to mitigate emittance growth effects and is the subject of this note.

\vspace{2.0mm}
\noindent{\bf Four types of bunch compression optics:} We now consider 4 styles of compression optics for use in the final compression stage: a simple 4-bend chicane; a minimal extension to provide CSR canceling- a 5-bend chicane; a double chicane with quadrupole optics between; and finally, a quadrupole (and sextupole) loaded arc or wiggler system.

\begin{enumerate}[topsep=4pt, itemsep=0.3ex, partopsep=0.3ex, parsep=0.3ex, label=\roman*.]
\item {\bf 4-bend chicane:} A classic 4-bend chicane is used for compression in (at least the later stages) many existing FEL facilities. It is simple to install and operate and has some nice features, such as automatic achromaticity to second-order, transversely. It is included here mainly for comparison purposes. This chicane system needs to be quite long at our 30 GeV design energy. The R56 of a chicane is also -ve, requiring it to ``fight'' the chirp induced in the rf structures of the final acceleration stage. There is no mechanism for canceling the CSR emittance growth (predominantly generated in the final 2 bend magnets).
\item {\bf 5-bend chicane:} Adding an additional bend to the chicane provides the minimum required additional free parameters to partially self-cancel emittance dilution due to CSR (e.g., see \cite{5-bend-chicane}). For our case, this design suffers from the problems outlined in the 4-bend chicane case (wrong sign R56, long footprint). Additionally, realistic simulations (such as \cite{5-bend-chicane}) show expected performance of CSR cancelation at about the $\sim 10\%$ level which for the most demanding cases (particle collider) would not be sufficient.
\item{\bf Dogleg or zig-zag chicane:} In order of increasing complexity, the next design to consider would be a double-chicane configuration (e.g., as considered in \cite{csr-compensation}). This directly cancels the CSR degradation in the first chicane by introducing a second chicane and carefully tuning the optics to arrange for the CSR cancellation, including a quadrupole-based lattice between chicane pairs. 
Although, from simulation studies, more performant in CSR cancelation than the 5-bend chicane design: For our high-energy requirement, this design suffers again from having the wrong R56 sign and more so from system length.
\item{\bf Arc or Wiggler compression lattice:} To achieve the desired +ve R56 compression in our final compressor design, we need a quadrupole-loaded lattice such as an arc or wiggler (depending on whether net-bending is a desired feature or not).

\begin{figure}[!htb]
\centering
   \includegraphics[width=0.5\columnwidth]{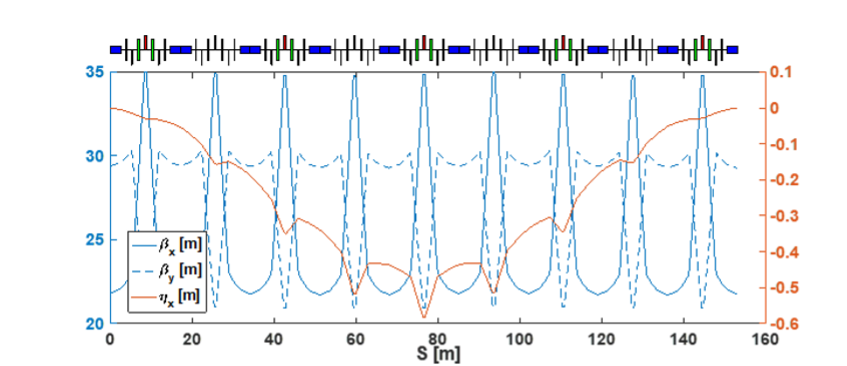}
   \caption{Simulation of Twiss parameters of 30 GeV arc compression lattice}
\label{fig:30GeV-arc-compression-simulations}
\end{figure}

To demonstrate the idea, we constructed a model of a wiggler system designed to perform the tasks of our final compressor using the tracking code Lucretia shown in Fig.\ref{fig:30GeV-arc-compression-simulations}. The design is a triplet-based arc (9 cells) with a total bend angle of 67 mrad @ 30 GeV. The arc provides +17mm of R56 in a length  of ~150m whilst keeping emittance growth to $<5\%$ due to incoherent SR in the bends. Sextupoles are included in the optics, as depicted in Fig.\ref{fig:30GeV-arc-compression-simulations}. Without CSR and with sextupoles deactivated, these optics are achromatic to second order.

\begin{figure}[!htb]
\centering
   \includegraphics[width=0.5\columnwidth]{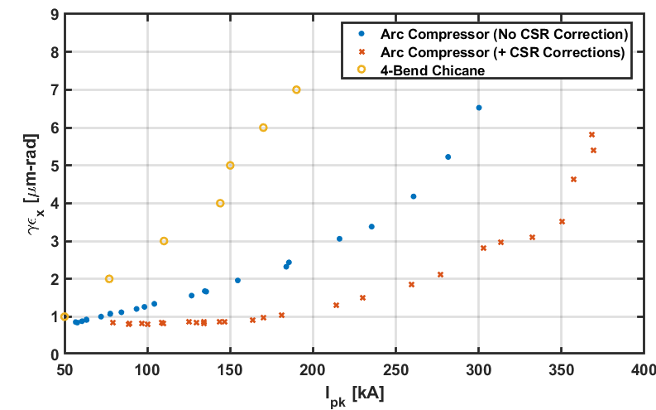}
   \caption{Results from particle tracking through the arc compressor, and 4-bend chicane for comparison. Slice emittance is shown as a function of peak current. The compression is controlled by the initial chirp of the beam and the optimization parameters.}
\label{fig:30GeV-arc-emittance-peak-current}
\end{figure}

Fig.\ref{fig:30GeV-arc-emittance-peak-current} shows the results of tracking simulations through these optics, with an initial optimization of sextupole strengths. The blue and red curves show the performance with/without sextupole optimization, respectively. The red curve represents the Pareto front output of the optimizer. The control variables used in the optimizer are the 18 sextupole strengths, in addition to 9 quadrupole strength knobs and fine control of the incoming energy chirp. The corresponding performance of a simple 4-bend chicane is shown for comparison. As seen, the arc compressor is a significant improvement over the chicane design, and optimization of sextupoles provide further improvements. The maximum compression for this design is limited at 375 kA, with 5 mm-mrad of slice emittance growth ($\rm \sim 15 mm-mrad$ full emittance growth including beam tails). Work is ongoing to further improve the emittance growth at higher peak currents.

\end{enumerate}

\vspace{2.0mm}
\subsection{Permanent Magnet Quadrupoles for high-field focusing \\ {\it contributed by G. Andonian}}
Excitation of the strongly electrostatic longitudinal fields in the crunch-in modes requires very dense drive beams within the core of the conducting nano-structures. Beam densities on the order of $\rm 10^{17}cm^{-3}$ to $\rm 10^{21} cm^{-3}$ are required to explore the different regimes of nano-plasmonic acceleration, which corresponds to transverse spot sizes on the sub-micron scale for intense beams. In order to produce conditions optimal for crunch-in mode formation, an appropriate focusing scheme must be implemented. Furthermore, in realistic experimental scenarios where regions are constrained by spatial limitations, the focusing system should be compact and robust.

\begin{figure}[htb]
\centering
   \includegraphics[width=0.4\columnwidth]{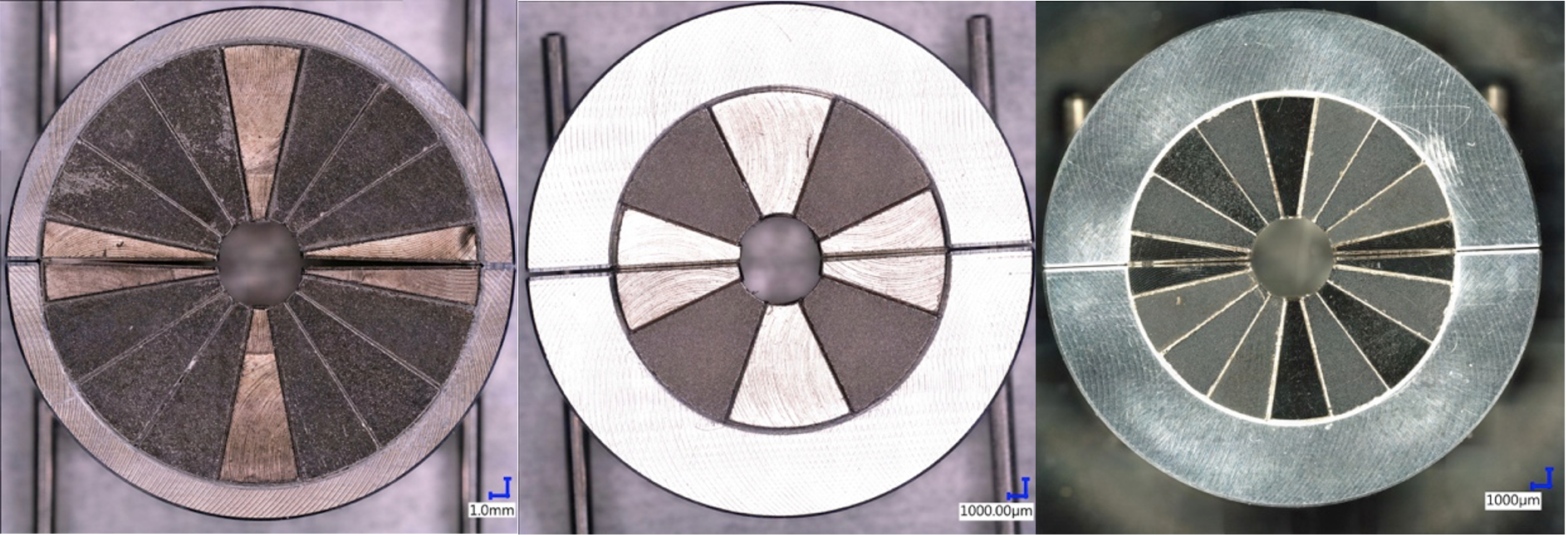}
   \caption{Examples of different permanent magnet quadrupoles in the Halbach geometry, produced for ultrafast electron microscopy at the Brookhaven National Laboratory. [Photographs courtesy of J. Penney.] }
\label{fig:PMQ-triplet}
\end{figure}

Permanent magnet quadrupoles are ideal candidates for such applications as they are able to provide very high gradients (in excess of 700 T/m \cite{adj-pmq}) in a compact form factor, without the need for external cooling that electromagnetic counterparts require. However, quadrupoles only offer focusing in one dimension, while defocusing the beam in the other. Hence, the simplest configuration of permanent magnet quadrupoles for overall focusing requires a triplet configuration. The triplet system is tunable by the relative reconfiguration of quadrupole locations, and higher order moments can be tuned out by magnetic shimming as needed.

Current applications of permanent magnet quadrupoles are varied across the accelerator physics landscape, including final focus systems for inverse Compton scattering \cite{adj-pmq}, beam-break up mitigation for extended energy exchange in wakefield interactions, and high quality imaging in ultrafast electron microscopy \cite{aberrations-tem}. Advances in material development and higher levels of sophistication in fabrication techniques also allow for control of higher order modes, such as in the Halbach configuration \cite{rare-earth-magnet}. Technological advances in permanent magnet quadrupole integration into existing beamlines, applied in-vacuum or out-of-vacuum with hybrid split-designs \cite{pmq-bnl-uem}, offer further flexibility for use within the strict constraints in nano-plasmonic acceleration.

\vspace{2.0mm}
\subsection{Electro-optic sampling (EOS) bunch-length diagnostics for intense beams \\ {\it contributed by G. Andonian}}
An advanced diagnostic technique of estimation of the relative position of a charged particle beam with the capability to resolve at the sub-micron scale uses electro-optic sampling (EOS). By measuring the change in the birefringence of a crystal subjected to collective fields of the beam makes it possible to estimate the bunch length and its distance from the crystal in addition to the possibility of estimating a few other beam properties. The change in birefringence depends upon the magnitude of the beam fields. This change in birefringence is measured by detecting the changes in the polarization of a laser propagating through a birefringent crystal. By autocorrelating the rotation of the polarization of the laser, it becomes possible to estimate the position of the beam. The resolution of EOS technique as a result depends upon the resolution of the polarization auto-correlator. EOS is a well-known technique that is currently being implemented in a novel configuration at the FACET facility at SLAC \cite{eos-diag}. These configurations can utilize multiple arms to not only estimate the position of the beam but also its bunch length and possibly also its longitudinal bunch profile.

\vspace{2.0mm}
\subsection{Advanced ionization beam-profile diagnostics for intense beams \\ {\it contributed by G. Andonian}}
The generation of the crunch-in modes in nano-plasmonic acceleration requires the delivery and transport of high intensity beams through small apertures. Traditional diagnostics for transverse beam profile and centroid for alignment would not be sufficient due to damage from the intense beams. Alternative methods, preferably non-destructive are required to parameterize drive and witness beams through nano-plasmonic structures.

An exciting new diagnostic tool for high intensity beams operating in a minimally intercepting topology is the gas sheet ionization monitor. The concept is based on the generation of a thin neutral gas sheet, or curtain, that is propagated perpendicular to the drive beam. As the beam passes the neutral gas sheet, the particles are ionized, leaving a footprint of the initial beam distribution. The ionization distribution is then imaged using a series of electrostatic lenses and a micro-channel plate. The ion distribution is then back-convolved using a reconstruction algorithm to determine the initial beam profile at the point of ionization. Such methods have been used at low energy \cite{gas-jet-bpm} and also proposed for high intensity beams at SLAC FACET \cite{high-intensity-bpm}. The technique is ripe for modern machine learning methods to extrapolate beam profile information where conventional techniques are not viable. 

\begin{figure}[!htb]
\centering
   \includegraphics[width=0.35\columnwidth]{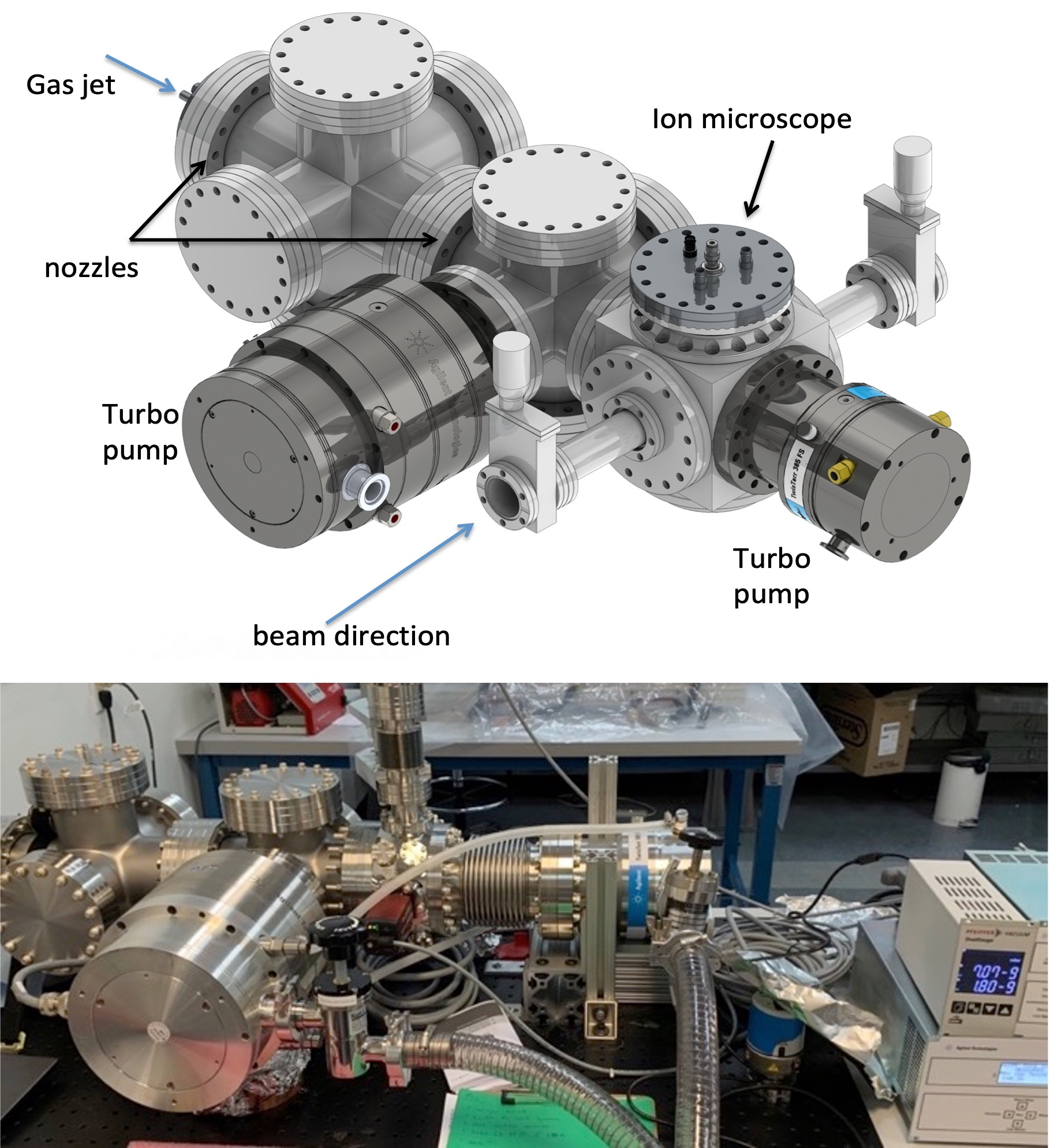}
   \caption{Gas film ionization beam profile monitor for dense beams.}
\label{fig:Gas-film-diag}
\end{figure}

\section{X-ray laser driven Carbon Nano-Tubes (CNT) \\ {\normalsize \it contributed by G. Mourou and T. Tajima}} 
\label{xray-laser-nanostructures}
\sectionmark{x-ray laser driven CNT}

The recent invention of the Thin Film Compression technique opens the way to introduce the availability of the single-cycled laser pulse and thus the Relativistic Compressed X-ray laser pulse \cite{xray-Mourou}. This innovation allows relativistically high-intensity X-ray laser pulse. This high-intensity X-ray laser pulse fits the need for X-ray laser driven nanostructured wakefield acceleration (xWFA) possibility \cite{Tajima-crystal-xray}. This technology and arrangements allow the level of accelerating gradient on the order of 100 TeV/m.

Although the energy level and time duration of x-ray or optical energy in the pre-pulse of the yet to be prototyped x-ray laser remains uncharacterized, it is critical to note that {\bf plasmonic processes are only relevant if this radiation energy ahead of the main x-ray pulse does not completely ablate the target and dismantle the ionic lattice}.

Another important path is nanostructures such as carbon nanotubes. The use of concavity of nanostructures such as CNT allows to eliminate collisions of the accelerated particles with the  ``background medium particles'' and to serve to conduct electrons with self-focusing force. This leads to the potential of on the order of TeV/ cm accelerating gradient possibility.  Nanostructures help also the ``confinement'' of the driving pulse of X-rays or charged particle beams. Furthermore, the introduction of such ultrahigh accelerating gradient opens up a path of non-luminosity paradigm of doing extreme high-energy physics, by stacking 1000 nano-fibers would lead to PeV over 10m.

Acceleration of muons (instead of electrons or hadrons) channeling between the planes in crystals or inside carbon nanotube (CNT) with high charge carrier density holds the promise of the maximum theoretical accelerating gradients of 1-10 TeV/m allowing envisioning of a compact 1 PeV linear crystal muon collider \cite{xray-CNT}. The choice of muons is beneficial because of small scattering on solid media electrons, absence of beamstrahlung effects at the IP, and continuous focusing while channeling in crystals, i.e., acceleration to final energy can be done in a single stage. Muon decay becomes practically irrelevant in such very fast acceleration gradients as muon lifetime quickly grows with energy. Initial luminosity analysis of such machines assumes a small number of muons per bunch O(1000), a small number of bunches O(100), high repetition rate O(1 MHz) and ultimately small sizes and overlap of the colliding beams O(1 Angstrom).

\section{Plasmonic accelerators: \\ Challenges, possibilities and efforts}
\label{challenges-prototype}
\sectionmark{Challenges, possibilities and ongoing efforts}

\subsection{Challenges and possibilities}

The challenges of the {\bf PV/m plasmonics} initiative are defined by its {\bf two distinct focal points}:
\begin{enumerate}[topsep=4pt, itemsep=-1mm, partopsep=0.3ex, parsep=0.3ex, label=\Roman*.]
 	\item in the near-term: direct access to frontier HEP using extreme plasmonic fields, or 
 	\item in the longer-term: ultra-high gradient accelerator modules for a future collider
\end{enumerate} 

\vspace{2.5mm}
\noindent{\bf Direct access to non-collider frontier HEP using PV/m plasmonics} \\
The physics case are already summarized above in Sec.\ref{non-collider-HEP}. The challenges that underlie the  have been identified are as follows:
\begin{itemize}[itemsep=-1mm]
\item computational modeling and material design
\item detailed modeling of physical processes behind plasmonic nanofocusing
\item in-depth understanding of nano-Wiggler mechanisms and its limits for coherent gamma-ray laser
\item specialized and unconventional diagnostics for ultra-solid beams of photons and particles
\item development of new diagnostic techniques to measure and experimentally characterize the modeled physics
\item design of short-term test cases of frontier HEP
\item developing a stepwise plan for attaining the limits of EM fields
\end{itemize}

\noindent{\bf Unforeseen applications of PV/m plasmonics} \\
The extreme fields of plasmonic modes help envision unforeseen possibilities. A few of these possibilities have been identified as part of the effort to educate the community on plasmonics and are listed here:
\begin{itemize}[itemsep=-1mm]
\item plasmonic accelerator can help realize crystalline beams
\item medical applications of ultra-compact accelerators for nano-surgery, or diagnostics inside the human body
\item beam-based inertial fusion and solving the world's energy problems
\end{itemize}

\noindent{\bf Plasmonic module for a future collider} \\ (contributed by T. Katsouleas and F. Zimmermann)
\begin{itemize}[itemsep=-1mm]
\item scaling of net power requirement for a collider-level machine with acceleration gradient
\item preliminary Integrated Design Study using plasmonic accelerator modules
\item beam dynamics questions - BBU, etc.
\item technological questions - staging and inter-stage alignment
\item nano-wiggler for coherent-gamma-ray laser for gamma-gamma collider
\item path to highest energy colliders
\item high energy accelerators for dark sector searches
\end{itemize}

\noindent{\bf Theoretical and computational modeling of plasmonic processes} \\
(with inputs from T. Katsouleas and D. Filipetto) \\

A comprehensive approach to modeling of relativistic and highly nonlinear plasmonic modes in conductive materials is required to appropriately incorporate the multi-scale and multi-physics nature of the underlying processes. These nonlinear plasmonic modes being strongly electrostatic significantly differ from the conventional optical plasmons and do not lend themselves to being modeled using purely electromagnetic codes. Optical plasmons are typically modeled using Finite-Difference-Time-Domain (FDTD) method where the perturbative electron oscillations are simply approximated using constitutive parameters. The FDTD approach with constitutive parameters cannot be used when the trajectories of collective electron oscillations attain amplitudes which are a significant fraction of the plasmonic wavelength and as a result requires a non-perturbative approach.

Our work has adopted the kinetic approach along with Particle-In-Cell (PIC) computational modeling of collective oscillations of the free electrons Fermi gas to account for the nonlinear characteristics of strongly electrostatic plasmons. It is noted that PIC methodology already utilizes the FDTD solver for calculation of electromagnetic fields but utilizes charge and current densities based upon particle tracking. This approach is utilizes the collisionless nature of relativistic oscillations of the Fermi gas. This approach does not assume any constitutive parameters as part of the initial conditions. However, the initialization and self-consistent evolution of the electron density implicitly accounts for most of the constitutive parameters. 

Moreover, as relativistic oscillations of the free electron Fermi gas have been experimentally observed to go beyond the Ohm's law, the long-term evolution of these oscillations still remains not fully accounted for by collisionless modeling approach. Specifically, conductivity which is a critical constitutive parameter based upon electron-ion collisions is not properly understood in the relativistic plasmonic regime. Therefore, longer-term processes resulting from the excitation of relativistic plasmonic modes are not fully incorporated in the kinetic modeling approach.

In consideration of this immense promise of PV/m plasmonics using nanomaterials, we call for support of the modeling community to engage with the our effort and understand several new challenges that are not part of conventional modeling tools.

Below we outline a few of the key challenges identified through our ongoing efforts and bring out the unique requirements our modeling effort:
\begin{itemize}[itemsep=-1mm]
\item Modeling the effect of relativistic energy gain of free electrons in a particle-in-a-box model
\item Understanding the effects of the energy density of plasmonic modes on the energy-band structure
\item Incorporating effects from atomistic modeling within the kinetic approach 
\item Devising an approach to handle collisions to determine the longer term effects of electrostatic plasmonic fields and account effects related to conductivity
\end{itemize}

\vspace{1.0mm}
\noindent{\bf Plasmonic materials research} \\ (contributed by D. Filipetto) \\

Material engineering is an essential ingredient of plasmonic accelerators. In such aspect, this research direction takes great benefit from the recent advances in nanofabrication techniques, which enable three-dimensional surface structuring of a plasmonic metal with single-digit nanometer precision. Also, the interface surface quality is of crucial importance in plasmonic accelerators, as any asperity can create local enhancement, field distortion, and be a source of energy depletion through radiative effects. The progress in electron beam lithography techniques, allows control of the material surface roughness at the Angstrom level \cite{plasmonic-emitters}.

\subsection{Ongoing and Future efforts}

Experimental effort to prototype PetaVolts per meter Plasmonics was proposed by the nano2WA collaboration at the SLAC FACET-II facility in 2020 \cite{nano2WA-plasmonic-expt}. This experimental proposal was evaluated by the program advisory committee (PAC) which prescribed the development of a short-term experimental plan \cite{PAC-close-out-report}. Several members of the PV/m Plasmonics community are active collaborators on the nano2WA experiment and are closely working towards a short-term experimental plan.

Besides the FACET-II effort there are currently several ongoing collaborative efforts to realize the potential of PV/m plasmonics by synergistically working with the plasma acceleration community. The quality and ultrashort nature of plasma accelerated beams can help further study plasmonic accelerators.

In view of the promise of PV/m plasmonics, in the future it may be a possible approach to request for a dedicated facility to study non-collider physics through directly access to test of HEP frontiers. Similar dedicated facilities have been proposed in the recent past for studies of non-collider physics and it may be possible to work together and expand the scope of those proposals.


\end{document}